# Non-invasive optical focusing inside strongly scattering media with linear fluorescence

Dayan Li,[1] Sujit Kumar Sahoo,[1,2] Huy Quoc Lam,[1,3] Dong Wang,[1,4] Cuong Dang, [1,*]

[1]*Centre for OptoElectronics and Biophotonics (COEB), School of Electrical and Electronic Engineering, The Photonics Institute (TPI), Nanyang Technological University Singapore, 50 Nanyang Avenue, Singapore 639798, Singapore*
[2]*School of Electrical Sciences, Indian Institute of Technology Goa, Goa 403401, India*
[3]*Temasek Laboratories @ Nanyang Technological University, 50 Nanyang Avenue, 639798, Singapore*
[4]*Key Laboratory of Advanced Transducers and Intelligent Control System, Ministry of Education, and Shanxi Province, College of Physics and Optoelectronics, Taiyuan University of Technology, Taiyuan 030024, China*
*\*Corresponding author: hcdang@ntu.edu.sg*

**Non-invasive optical focusing inside scattering media is still a big challenge because inhomogeneous media scatter both incoming photons for focusing and outgoing photons for observation. Various approaches, utilizing non-linear fluorescence or ultrasound, have been reported to address this difficulty. However, implementation of these methods is complicated and highly expensive, as ultrafast laser systems or photo-acoustic equipment must be employed. Here, we demonstrate a wavefront shaping technique to achieve non-invasive focusing (NiF) inside scattering media using only a linear fluorescent signal. Contrast and mean of incoherent speckles, produced by the linear fluorescence, are utilized as feedback signals to optimize the input wavefront. While increasing speckle contrast makes the focus tighter, and increasing the speckle mean enhances the intensity, fine-tuning the contribution of these two factors in our two-step optimization is essential. An optimal wavefront is found to achieve simultaneously both a micrometer focal spot size (down to 20 μm diameter) and high intensity (more than a 100-fold enhancement) inside the scattering media. Our method promises a new route in life science towards focusing, imaging or manipulating deep into biological tissues with linear fluorescent agents.**

## Introduction

Optical scattering poses major limits with respect to imaging resolution and depth when modern microscopic techniques are applied to complex media such as biological tissues. The inhomogeneous refractive index of complex media strongly perturbs light propagation and thus prohibits the creation of diffraction-limited focus. While appearing random, light scattering in static media is a deterministic process, and therefore its effect can be 'canceled' or reversed [1, 2]. One can implement wavefront shaping techniques to achieve focusing through scattering media with direct feedback from internal sensors, i.e. invasive approaches [1, 3, 4]. However, in scenarios where direct access to the target plane is not possible [5], the problem becomes much more challenging. Such non-invasive approaches are much more attractive for many medical and biological applications.

Various non-invasive methods (see ref. 6 and references therein) have been proposed to tackle light scattering issues and then focus light into complex media. One can use acoustic waves to eliminate the scattering effect and then non-invasively access the target plane in a photoacoustic feedback approach. The acoustically focusing spot defined by the resolution of the ultrasound wave is the 'guide-star' for wavefront shaping [7-10]. Fully optical approaches usually employ non-linear fluorescence [11, 12] (e.g. the two-photon or multi-photon effect). Because photo-acoustic coupling or non-linear excitation is involved, these methods are subject to the drawbacks of complication and cost in implementation. Using time-gating in coherent systems also helps non-invasive wavefront corrections [13]. When combined with random scattering matrix [14], time gating shows effectiveness in non-invasive focusing [15] or imaging [16, 17] deep inside scattering media. However, these methods [15-17] are preferably being applied to discrete objects; a limitation which is related to the feasibility of implementing 'iterative time reversal' [18, 19]. Recently, non-invasive focusing using speckle correlations has been proposed [20]. Because this approach is based on the optical memory effect, it is only applicable to relative thin samples and the depth of focus is shallow.

Here, utilizing linear fluorescence as an optical feedback signal, we demonstrate an iterative wavefront shaping technique that is able to non-invasively focus light onto an extended object behind strongly scattering media. The statistical properties of the fluorescence-generated speckle pattern, such as its contrast and mean, are shown respectively related to the focal spot size and intensity enhancement in creating a focus. Because the speckle pattern at the imaging sensor is the incoherent summation of multiple speckle patterns generated by each individual fluorescent source, more sources will reduce the speckle contrast. Therefore, maximizing speckle contrast will concentrate the excitation light onto a single fluorescent source. On the other hand, mean intensity of speckle pattern relates to the amount of excitation light that falls onto the fluorescent sample. By increasing speckle variance (the product of mean intensity and speckle contrast) in optimization, Gigan and co-workers achieved non-invasive focusing on scattered fluorescent beads [21]. In fact, the approach showed the dominance of mean intensity in optimization, and therefore, quickly achieved a loose focus region when the number of discrete fluorescent bead increased. In contrast to this, we implement a two-step optimization approach with controllable contributions of mean intensity and speckle contrast. We start with a mean dominant optimization, followed by a contrast dominant optimization to achieve a tight focal spot with high intensity on a continuously extended object. An optimized wavefront is found to create a focal spot of 20 μm diameter behind the scattering media, while its intensity is enhanced up to 100-fold. Our results demonstrate a major step to control light through strongly scattering media, promising various applications in life science such as optical surgery, drug delivery, and imaging.

## Principle

A schematic of the non-invasive focusing inside scattering media, as well as a numerical example, are presented in Fig. 1. Input laser light $E_i$ is scattered before hitting a fluorescent object inside the scattering media. Due to multiple scattering, only a small portion of $E_i$ can reach to the depth of the object. We illustrate this situation in Fig. 1a, where $E_e$ denotes fluorescence excitation light, $E_f$ is the portion of light scattered forward but not reaching to the object and $E_b$ is backward scattered light. The purpose of our non-invasive focusing is to minimize $E_f$ and $E_b$, thus $E_e$ is maximized, and to further squeeze $E_e$ into a single focal spot. The fluorescent sample absorbs the excitation light coming to it then emits fluorescent light at longer wavelength. The imaging sensor detects the scattered fluorescent light after passing through the scattering media. To achieve a focusing spot on the sample, the input wavefront of $E_i$ must be shaped to overcome the effect of multiple scattering for excitation light while the fluorescent speckle pattern at the imaging sensor is the feedback signal. Here, we propose to use wavefront shaping based on genetic algorithm (GA) to achieve the goal, while other optimizing algorithms are also available[22].

On the object plane, assuming that a fluorescent object can be sampled into $K$ sources. For the single $k^{th}$ source, it emits fluorescent light and, after multiple scattering, generates a speckle pattern $S_k(x)$ at position $x$ on the detector plane. Then $S_k(x)$ is point spread function (PSF) describing fluorescent light propagating from the $k^{th}$ source to detector plane. Unlike the approach relied on memory effect where the PSF is shift-invariant [23, 24], our PSF of $S_k(x)$ can be source dependent, i.e. no memory effect requirement. This can significantly enhance the depth of focusing. The total intensity $I(x)$ at the detector is an incoherent sum of all the speckles generated by each individual source. As a result, $I(x)$ at detector is given by:

$$I(x) = \beta \sum_{k=1}^{K} \left| E_e^{(k)} \right|^2 S_k(x),  \quad (1)$$

where $E_e^{(k)}$ is the excitation light field that illuminates the $k^{th}$ source. Because it is a linear fluorescence, the response of fluorescence is proportional to excitation intensity rather than the field of illuminating light. $\beta$ denotes fluorescence's photon conversion efficiency.

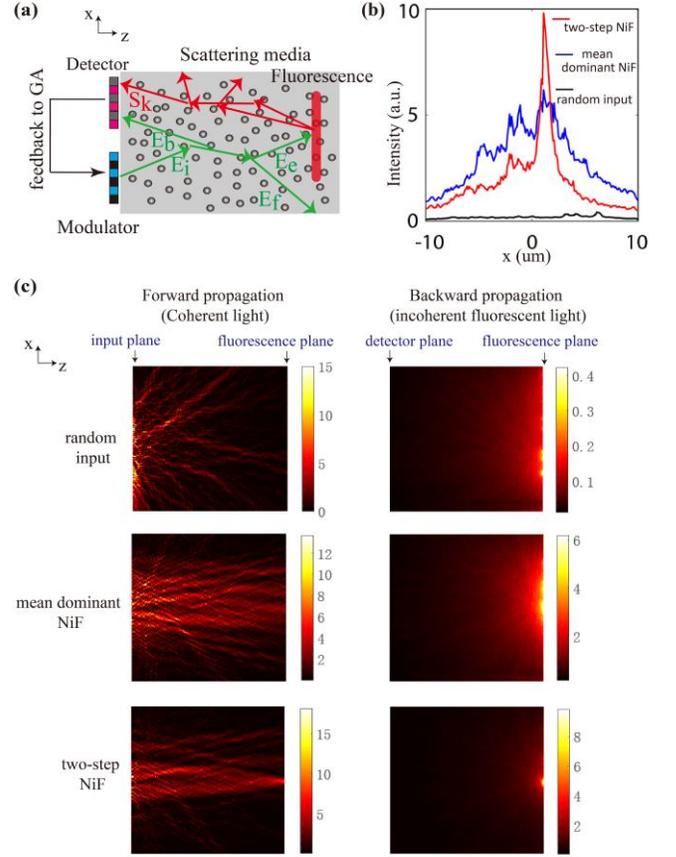

Fig. 1. **Non-invasive optical focusing inside scattering media: concept and numerical demonstration.** (a) Schematic of non-invasive focusing (NiF) inside scattering media by wavefront shaping. $E_i$ denotes input coherent light; $E_f$, forward scattered light not reaching to the object; $E_b$, backward scattered light; $E_e$, forward scattered light reaching to fluorescent object; $S_k$ denotes speckle at detector generated from single $k^{th}$ fluorescent emitting source. Input wavefront is optimized by a GA with feedback from speckle statistics at the detector. (b) Illustrating results of non-invasive focusing from simulation: intensity profiles of fluorescent light at fluorescence plane corresponding to c are shown. (c) Wave propagation simulated in x-z plane, including both forward excitation light and backward fluorescent emission, for random input wavefront (top) and optimized input wavefronts from mean dominant NiF (middle) and two-step NiF (bottom), respectively.

From Eq. (1) and theory of speckle statistics [25], we know that contrast of $I(x)$ pattern, C($I$), is inversely proportional to $\sqrt{K}$. As a result, if one can optimize the input wavefront such that fewer fluorescent areas are excited, the contrast C($I$) will be increased. In principle, when only a single source is excited, i.e. creating a focus, C($I$) reaches its maximum value. We also see from Eq. (1) that total intensity of $I(x)$ is linearly proportional to the mean value of illuminating intensity on the fluorescent object. In order to create a focus with higher intensity enhancement, one also needs to optimize for M($I$), the mean value of $I(x)$.

Here, we can define a flexible cost function, $F(I) = \alpha * C(I) + M(I)$, as the feedback signal for GA, where $\alpha$ is a constant controlled by users. When $\alpha * C(I) \ll M(I)$, GA implements a mean dominant NiF. Similarly, when $\alpha * C(I) \gg M(I)$, a contrast dominant NiF will be carried out. By controlling $\alpha$, we can fine-tune the optimization approach in a broad range from mean dominant NiF to contrast dominant NiF to achieve the goal. Mean dominant NiF tends to increase amount of excitation light onto the fluorescence area, which might lead to a loose focusing spot or multiple spots (loose focusing). On the other hand, subsequent contrast dominant NiF promotes a shrinking size of excitation spot on the fluorescence sample, leading to a tighter focus spot; however, intensity enhancement is ultimately limited.

As a numerical example, the system depicted in Fig. 1a is simulated. Thanks to μ-diff [26], an open-source Matlab toolbox for solving the multiple scattering problem, scattered light fields inside two dimensional (x-z plane) scattering media can be easily calculated. The forward propagation of coherent excitation light and backward propagation of fluorescent emission light, as well as GA based wavefront shaping, are all simulated and integrated together, so that non-invasive focusing is numerically investigated. Details about the simulation method can be found in the supplementary material [27].

Fig. 1b shows the intensity profiles at the fluorescent object plane for different input wavefronts. We see that at the initial condition, when a random wavefront is illuminating the scattering media, fluorescent intensity is low and almost flat, implying the multiple scattering effect imposed to the input light. Fig. 1c top illustrates this situation. We see that input light is scattered out of the scattering media and only a small portion of light can reach the fluorescent object. In return, only small amount of fluorescent emission is observed. In Fig.1 c top, we also observed obvious light interference arising from coherent waves propagating in the scattering media, while the incoherent fluorescent emission light does not have such effect. After wavefront optimization with mean dominant cost function, we indeed see a significant increase of fluorescent intensity (blue curve in Fig. 1b). This corresponds to Fig. 1c middle. As can be seen, a very large portion of input light is guided onto the florescent object, which implies that $E_e$ is enhanced by mean dominant NiF. However, at this stage, only a loose focus is obtained; which is similar to optimization with speckle variance as the cost function [21]. Continuing the wavefront optimization process, but now switching to contrast dominant NiF, finally we obtain a tighter focus, with intensity at the focal spot further increased (Fig. 1b red curve). From Fig. 1c bottom, we clearly see that more $E_e$ is squeezed onto a reduced fluorescent area.

## Experimental realization of non-invasive focusing inside scattering media

Fig. 2 presents the experimental setup for non-invasive focusing inside scattering media by feedback-based wavefront shaping. A CW laser (λ = 532 nm, Thorlabs TCLDM9) is expanded onto a phase-only spatial light modulator (SLM, Holoeye Photonics, Pluto-2). Laser light covers 640 by 640 SLM pixels in total, and we segment those pixels into 64 by 64 controllable modes for wavefront shaping. SLM is imaged onto the front surface of the scattering media, since conjugated adaptive optics (AO) has a much larger field of view in wavefront corrections than pupil AO [28]. Lens $L_3$ and objective lens $MO_1$ (Nikon S Plan Fluor, ELWD, 40X, NA 0.6) constitute the imaging system.

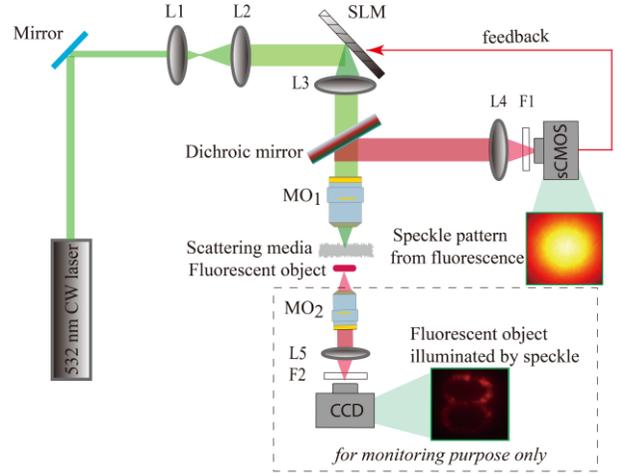

Fig. 2. **Experimental setup.** L1-L5: lens, $MO_1$ and $MO_2$: microscope objective, SLM: spatial light modulator, $F_1$ and $F_2$: band-pass filters, which are used to reject the excitation green laser and only pass the fluorescent emissions. A fluorescent object is placed beneath a scattering media; its fluorescent emission is collected in a reflection configuration by $MO_1$, $L_4$ and a sCMOS camera. Inset shows a typical incoherent speckle pattern at the sCMOS camera. SLM is imaged onto the scattering media by $MO_1$ and $L_3$, implementing a scattering media conjugated wavefront shaping approach. $MO_2$ and $L_5$ constitute an imaging system to "see" the object, which serves the monitoring purpose only.

We prepared our fluorescent object by drop-casting red quantum dot (QD) solution, emitting at 630 nm with full width at half maximum (FWHM) of 30 nm, on a designed chromium mask (see Fig. 2 inset). The samples are typical numbers with 375 μm in height, 256 μm in width, and 32 μm linewidth. A short-pass dichroic filter (Edmund, cut-off wavelength 550 nm) is used to facilitate a reflecting detection configuration. The fluorescent light is collected by $MO_1$ and $L_4$, recorded by a sCOMS camera from Andor (Neo 5.5). $F_1$ and $F_2$ are band-pass filters (Semrock, FF01-625/26-25), which reject the excitation green light and pass fluorescent emissions. Another objective lens (Edmund, 20X, NA 0.4) $MO_2$, together with a lens $L_5$ and a camera (Thorlabs, DCC1645C), is used to monitor the illumination and record non-invasive focusing results for reference purposes only. In the experiment, we used cascaded two ground glass diffusers (Thorlabs, N-BK7, 120 Grit) as scattering media.

Fig. 3 presents the focus development during the iterations of wavefront shaping, our proof of concept demonstration. During the wavefront optimization process, we first carry out mean dominant NiF, by choosing the $\alpha$ value such that the cost function $F(I) = \alpha * C(I) + M(I)$ is dominated by mean value. For example, here we set $\alpha = 170$, so that the initial ratio of $\alpha * C(I)$ to $M(I)$ is 0.1; see Fig. 3(a) left axis. It is observed that the cost function value continues to increase until iteration 320, when it then reaches saturation. If we compare the object image at iteration 320 (Fig. 3d) to that at initial condition (Fig. 3c), it is obvious that a considerable amount of light has been directed to the fluorescent object. By summing up the intensity values for the pixels on the object area, we estimate that fluorescent light intensity has been enhanced by 12.12 times after these iterations. However, despite remarkable intensity enhancement, the optimization only achieves a slight focusing effect. This result is understandable as enhancing the mean of a linear signal is equivalent to increasing the excitation light on the fluorescent sample.

In Fig. 3b, we also plot the focal spot size and intensity enhancement factor at the focal point over different optimization iterations. The focal spot size is calculated by counting the number of pixels that have intensity values greater than 10% of maximum intensity at each object image. The calculated spot size at each iteration is normalized by the spot size calculated from the object image at initial condition. As for the intensity enhancement, the value is calculated as the ratio of the focal point intensity at each iteration to the corresponding value at the initial condition. From Fig. 3b, we see that over the first 320 iterations, while intensity at the focal point is continuingly enhanced (~45-fold), the focal spot size is hardly reduced. This observation agrees well with our previous observation in simulation and our intuition that mean dominant NiF mainly optimizes intensity at fluorescent object. Some small contribution from speckle contrast in cost function is insufficient to create a tight focal spot.

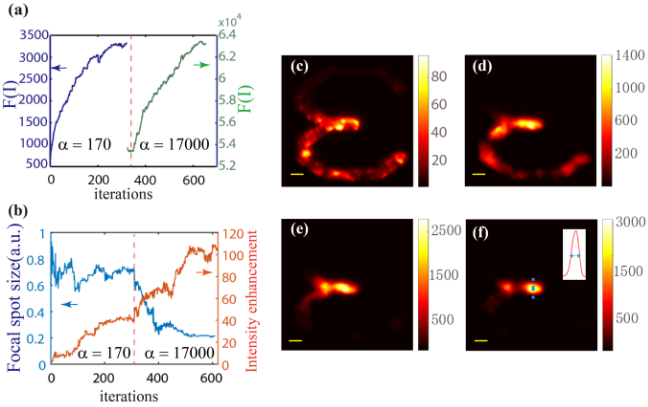

Fig. 3. **Experimental results of focus development.** (a) Plot of the cost function value at each iteration during optimizations. Left axis, mean dominant NiF; right axis, contrast dominant NiF. (b) Plot of focal spot size and intensity enhancement at position of focus over iterations. Images on object plane at some typical iterations: (c) before optimization, (d) at iteration 320, (e) at iteration 530, (f) at iteration 620. The intensity in the captured images (8 bits) is normalized with the exposure time to display. Inset: intensity profile of focal spot through dashed blue line in (f), which shows that the focus has a nice Gaussian shape. Scale bar: 32 μm.

After gaining a strong fluorescent signal with mean dominant NiF, we switch the optimization to contrast dominant NiF at the basis of iteration 320. Here, we set $\alpha = 17000$, where now the ratio of α*C(I) to M(I) is 60. From Fig. 3a right axis, we see that the cost function value increases during optimizations and reaches saturation after about 620 iterations. Interestingly, the blue curve in Fig. 3b and object images in Fig. 3e and Fig. 3f show that a focal spot is gradually created during the contrast dominant NiF process. For example, comparing the object image at iteration 520 (Fig. 3e) to that at 320 (Fig. 3d), we visually observe the shrinking of object size. More importantly, we see from Fig. 3b that the intensity at the focal point continues increasing, while its size is reducing by contrast dominant NiF. Finally, at iteration 620 (Fig. 3f), a focal spot is created with saturation of the contrast dominant cost function. Thanks to our two-step optimization, intensity at focal spot has achieved about 100-fold enhancement compared to the initial object (red curve in Fig. 3b). The focal spot size is Gaussian with measured FWHM of 20 μm, as shown in the inset on Fig. 3f.

Now, we examine the speckle evolution over the iterations of wavefront shaping as presented in Fig. 4. The first observation is that the speckle area is shrinking over iterations (Fig. 4a-c). This is good indication that the fluorescent area at the object plane is also reducing. During the first stage of optimization, i.e. mean dominant NiF, the speckle's mean intensity increases as expected. Also, the standard deviation, which is essentially dominated by speckle mean, increases as well. After switching to contrast dominant NiF, both mean and standard deviation decreases (Fig. 4d and f). More importantly, the contrast of speckle increases monotonically, and certainly, the increasing rate is much more prominent when contrast dominant NiF is implemented (Fig. 4e). This implies that the focal spot starts to be created at the very first iterations, but it is not completely successful until contrast dominant NiF is applied. These observations at speckle statistics agree well with the object evolution described in Fig. 3 above. In Fig. 4f, we also plotted the intensity value at the focus over optimization iterations. We noted that, although mean intensity of speckle patterns is reduced by contrast dominant NiF, the intensity value at the focal spot is continually increasing during optimizations.

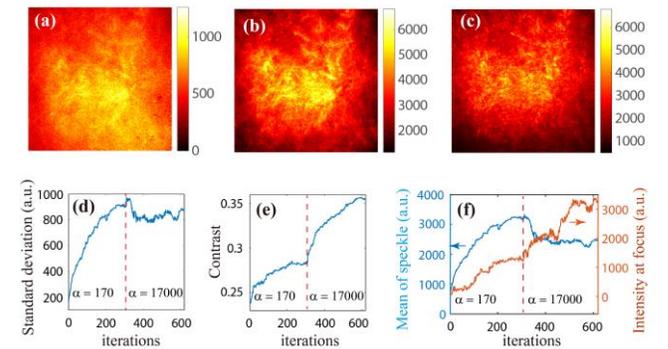

Fig. 4. **Speckle evolution during non-invasive focusing.** Speckle patterns at typical iterations: (a) before optimization, (b) at iteration 320, (c) at iteration 620. Speckle statistics including standard deviation (d) and contrast (e) over iterations. (f) Mean of speckle and intensity at focal spot over iterations. Exposure time of camera is 0.1s.

It is obvious that the speckle mean over optimizing iterations correlates to the excitation intensity at fluorescent object area. As mentioned before, light intensity at sample has been enhanced by 12.12 times after first 320 iterations using mean dominant NiF. Correspondingly, the mean of speckle patterns increases roughly from 500 to 3300 (Fig. 4f blue curve). The role of the subsequent contrast dominant NiF is to shrink the extended focus after mean dominant NiF into a tight focus while the increase of total fluorescent intensity is less emphasized. The actual behavior is even opposite with decreasing fluorescent intensity. If we calculate the light intensity at the object area, intensity is reduced 22.6% at the iteration 620 (Fig. 3f), compared to that at iteration 320 (Fig. 3d), the end of mean dominant NiF. However, while some of the excitation light goes out of object during contrast dominant NiF, some of the excitation light continues to be pulled into the focal spot, increasing its intensity (Fig. 4f red curve). This reduction of intensity at object area, while creating a tighter focal spot, is somewhat related to the essential limit of wavefront shaping, as initially proposed in Ref. [29]. There is a trade-off between focal spot size and intensity enhancement for wavefront shaping based optical focusing through scattering media [29]. This trade-off arises from the fact that each scattering event will redirect light within a cone of divergence angle, rather than distribute light equally over a solid angle of $2\pi$. The radially degrading intensity distribution, which is a typical scenario for scattering media, has a direct impact on both the focal intensity and the effective numerical aperture. As a result, even with wavefront shaping, excitation light at transversal locations further away from the focal position will have less or zero contributions on intensity enhancement. The larger the focus spot is, the more excitation light could be collected on that focal area through wavefront corrections; thus intensity enhancement is larger [29].

## Discussions

One needs to control the balance between mean (intensity) and contrast (focal spot size) by choosing an appropriate $\alpha$ in optimizations. If standard deviation serves as a cost function in the optimizations, it can be considered as single-step NiF with mean dominance, and our experimental results show a very loose focus (supplementary Fig. S3 [27]), similar to that in Ref. [21]. One can get a tighter focus with larger $\alpha$ coefficient (i.e. more weight on contrast), but then the ultimate intensity at focal point is reduced. One can fine-tune the coefficient $\alpha$, or even implement an adaptive approach, to adjust this coefficient continuously during optimization to achieve a better result. It is also advisable to start with mean dominant NiF to bring more excitation laser light into the fluorescent area; thus enhancing the fluorescent signal. There is a high chance that we achieve very little, or even no fluorescent speckle, when we do contrast-dominant NiF first. The speckle pattern is simply a camera's noise image, which also has high contrast. We would like to note that the non-invasive approach has no control of focusing position. The optimization process tends to focus on a bright spot, which is associated with thicker or more efficient fluorescent materials.

Unlike many other methods [15-17] that rely on discriminating ballistic photons from multiply-scattered photons for imaging or focusing purpose, our proposed technique utilizes the statistics of the speckle pattern that is generated from multiple scattering. As a result, our technique to some extent is immune to multiple scattering. It works until to the depth where multiple scattering is so strong that the speckle pattern is independent from the object, i.e. the wash-out effect. Here, for a proof-of-concept demonstration only, we show that our technique is capable to non-invasively focus light through the optical scattering thickness formed by cascading two frosted glass diffusers. The capability of the technique with respect to focusing depth will be left for future exploration.

## Conclusion

We have proposed a non-invasive wavefront shaping method which is able to focus light inside scattering media with the aid of linear fluorescent signal. In the optimization, contrast and mean of fluorescence-generated speckles, are respectively related to focal spot size and intensity enhancement. Optical focus inside scattering media can be non-invasively created as long as contributions of contrast and mean are appropriately controlled in optimizations. Our two-step NiF approach first emphasizes on increasing the excitation intensity at the fluorescent area by mean dominant NiF then reducing the focal spot size by contrast dominant NiF. Micro-scale spot size with hundred-fold intensity enhancement is achieved through strongly scattering media non-invasively. With our proposed technique, the fluorescent speckle pattern from the focusing spot is essentially a PSF. This non-invasively measured PSF allows us to do imaging around the focusing spot by deconvolution [24-25, 30-35], or move the focus spot to nearby positions within the memory effect region as presented in our supplemental Fig. S2 and S3 [27]. Our non-invasive focusing through scattering media with linear fluorescent feedback could provide a simple and useful tool for various applications in life science such as optical surgery, drug delivery, imaging or light therapy.

**Funding.** Ministry of Education - Singapore (MOE2019-T1-002-087).

**Acknowledgment**. We would like to thank the financial support from Singapore Ministry of Education through AcRF Tier1 grant MOE2019-T1-002-087. We also thank Miao Qi and Prof. Wei Lei for lending us the CW laser (Thorlabs TCLDM9) to finish the experiments, and Dr. P. A. SURMAN for careful reading of the manuscript.

**Disclosures.** The authors declare no conflicts of interest.

See Supplement 1 for supporting content.

# Non-invasive optical focusing inside strongly scattering media with linear fluorescence: supplementary material

DAYAN LI,[1] SUJIT KUMAR SAHOO,[1,2] HUY QUOC LAM,[1,3] DONG WANG,[1,4] CUONG DANG, [1,*]

[1]*Centre for OptoElectronics and Biophotonics (COEB), School of Electrical and Electronic Engineering, The Photonics Institute (TPI), Nanyang Technological University Singapore, 50 Nanyang Avenue, Singapore 639798, Singapore*
[2]*School of Electrical Sciences, Indian Institute of Technology Goa, Goa 403401, India*
[3]*Temasek Laboratories @ Nanyang Technological University, 50 Nanyang Avenue, 639798, Singapore*
[4]*Key Laboratory of Advanced Transducers and Intelligent Control System, Ministry of Education, and Shanxi Province, College of Physics and Optoelectronics, Taiyuan University of Technology, Taiyuan 030024, China*
*\*Corresponding author: hcdang@ntu.edu.sg*

**This document provides supplementary information to "Non-invasive optical focusing inside strongly scattering media with linear fluorescence," *XXX* volume, first page (year). This document includes: 1. Methods on numerical simulation of non-invasive focusing (NiF) inside scattering media, 2. Non-invasive imaging inside scattering media by deconvolution, 3. Non-invasively focusing light at arbitrary positions with the aid of the optical memory effect, 4. Non-invasive focusing using standard deviation as cost function.**

## 1. Numerical simulation method of NiF.

In this section, we describe in detail how we simulate non-invasive optical focusing inside two-dimensional scattering media. The schematic of the simulated system is shown in Fig. 1a of the main manuscript. Our basic tool is μ-diff [1], from which we construct a scattering medium in x-z plane and calculate multiply scattered light fields inside the medium. In our simulation, the size of the medium is 10 μm by 10 μm, with 100 randomly distributed disks. These scattering disks have random lengths in radius and constant reflective index of 14.14. The space in the media without occupying disks has a reflective index of 1, thus light will be scattered when hitting the disks.

We simulate the scattered light fields $T_m^{out}$ inside the media with point source illumination (*out* corresponds to the plane for observation and *m* is discrete position along x on that plane). We scan the illuminating point source along the x direction so that random scattering matrix $T_{mn}^{out}$ of the media is constructed (*n* is the discrete position along *x* on illuminating input plane) [2]. Because each scanned point source represents an independent input mode, output fields resulting from any illumination will be the linear combinations of the responses from each input mode, that is:

$$E_m^{out} = \sum_{n=1}^{N} T_{mn}^{out} E_n^{in} . \quad (S1)$$

With Eq. (S1), the response of wavefront shaping can be easily simulated. In this work, we assume the input illumination has a Gaussian amplitude $A$ with phase $\varphi(n)$,

$$E_n^{in} = A exp[j\varphi(n)] . \quad (S2)$$

The goal of GA based wave-front shaping is to find the optimized $\varphi(n)$ such that light is focused onto fluorescent target despite the multiple scattering effect presented by the random media.

For our non-invasive focusing simulation, $E_m^{out}$ is at the fluorescent plane. As explained in the principle in main manuscript, the response of fluorescence to excitation light is linear with intensity. The intensity at the detector is a linear combination of fluorescent intensity responses from individual sources, duplicating Eq. (1) of main manuscript for numerical simulation, as:

$$I_l = \beta \sum_{k=1}^{K} |E_k^{(out)}|^2 S_{lk} , \quad (S3)$$

where $S_{lk}$ is a matrix representing point spread functions (PSF). $S_{lk}$ in simulation can be constructed in same way as $T_m^{out}$, but this time the input plane is the fluorescent plane and output plane is the detector plane. By scanning point source illuminations, we can construct another random scattering matrix $T_{lk}^{(back)}$, and $S_{lk}$ is given by:

$$S_{lk} = |T_{lk}^{(back)}|^2 . \quad (S4)$$

This simple absolute square operation transforms the system from coherent one to incoherent one, which is needed for an optical system when it is illuminated with incoherent source [3]. $S_{lk}$ contains information regarding to how the light propagates in the scattering medium when illumination is incoherent point sources. The light scattering effect and energy decay caused by multiple scattering are the most important factors in our simulation, for which Eq. (S1 and S4) has taken into account.

Once the forward and backward propagations of light in the medium have been simulated, the resulting data can be sent to conventional GA, and numerical simulation of NiF is achieved.

## 2. Non-invasive imaging inside scattering media by deconvolution

In this section, we show non-invasive imaging inside scattering media can be easily achieved by a simple deconvolution operation. The experimental setup is exactly Fig. 2 in the main manuscript, only the scattering media is changed to a single ground glass diffuser to ensure that optical memory effect is presented. Since a focal spot inside scattering media has been created non-invasively, we can record the corresponding speckle pattern (Fig. S1b). This speckle pattern is the PSF of the system. Thanks to optical memory effect, the system is spatially shift invariant, that is, the PSF is only spatially shifted but not structurally changed if focal spot on object is emitting as point source at other places. As a result, with a single PSF de-convolved with object intensity at detector plane (Fig. S1a), the whole fluorescent object illuminated by laser speckles can be reconstructed (Fig. S1c).

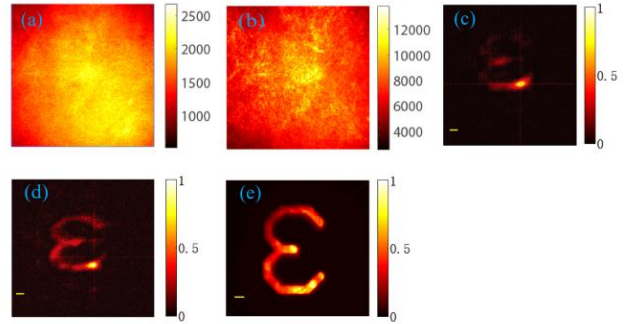

Fig. S1. **Non-invasive imaging inside scattering media by deconvolution.** (a) Object's speckle pattern at the detector plane. (b) PSF. (c) Reconstruction by deconvolution. (d) Average over many realizations. (e) Ground truth image. Scale bar: 32 μm.

Almost all the object structures have been retrieved with a single deconvolution operation. However, the fluorescent sample is speckle-illuminated by excitation laser through scattering media; therefore the deconvolution result is full of speckle. In order to remove speckles, we record many different realizations of the object's speckle patterns, by using many different random wavefronts as inputs. By de-convolving with the same PSF and averaging the reconstructed images together, we indeed obtained an object image (Fig. S1d) with better quality. In Fig. S1e we show the ground truth image for comparison.

## 3. Non-invasively focusing light at arbitrary positions.

With the aid of the optical memory effect, we can non-invasively focus light at arbitrary positions inside scattering media. As demonstrated above, object image can be retrieved after non-invasively focusing light inside scattering media. This enables us to "see through" scattering media. As a result, we can pick up arbitrarily any target position on object to create a new focus. The procedure is simple; at each iteration during wave-front shaping an object's image is reconstructed by deconvolution, and the intensity value at the target position is used as the cost function. In this way, a new focus will be created, despite the fact that we do not have a direct access to the target position. Fig. S2 shows the experimental results where we have arbitrarily focused light at five different positions on the object.

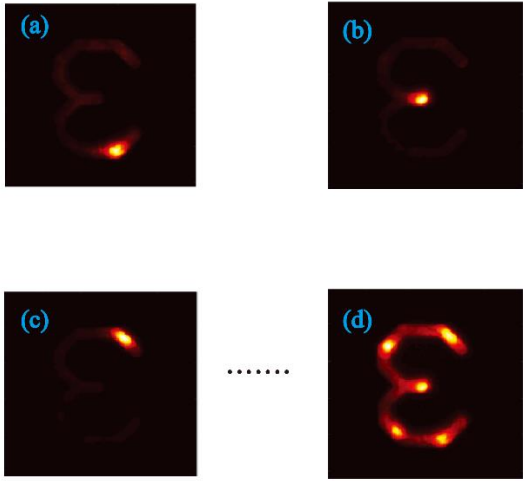

Fig. S2. **Focusing light at arbitrary positions.** (a-c) Focal spot at 3 different positions. (d) Many focal spots are combined (a superimposed image).

### 4. Non-invasive focusing using standard deviation as cost function.

In this section, we show the experiment when using standard deviation of speckle as the cost function for a continuously extended object. Standard deviation is mathematically the product of mean and contrast, and for a typical speckle pattern mean is far greater than contrast. As a result, mean will dominate the contributions in creating focus. From Fig. S3 (a) and (c), we see that the evolvement of standard deviation and mean almost follow the same trend during optimizations, while contrast is saturated at some point Fig. S3 (b). As a result, only loose focus can be created, as seen in Fig. S3 (e). Despite much energy having been brought onto the object area, a focal spot is hardly created. These experimental results clearly demonstrate that optimization only by standard deviation is simply mean dominant NiF as we show in the main manuscript.

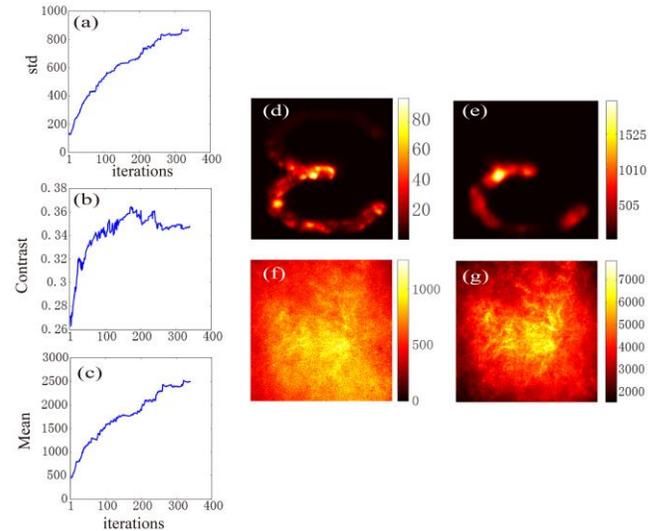

Fig. S3. **Non-invasive focusing using standard deviation as the cost function.** (a) Standard deviation as the cost function is used in GA-based wave-front shaping. Evolvement of contrast (b) and mean (c) of speckle patterns during optimizations. Object images before (d) and after (e) the optimizations, and the corresponding speckle patterns (f) and (g). The intensity images in (d) and (e) (8 bits) is normalized with the exposure time to display.